\documentclass[prb,preprint,showpacs]{revtex4} 

\usepackage{graphicx}
\usepackage[]{hyperref} 	

\newcommand{\CaW}{\mbox{$\mbox{CaWO}_4$}}
\newcommand{\alo}{\mbox{$\mbox{Al}_2\mbox{O}_{3}$}}

\newcommand{\CoF}{\mbox{${}^{57}\mbox{Co}$}}
\newcommand{\FeF}{\mbox{${}^{55}\mbox{Fe}$}}
\newcommand{\cmsq}{\mbox{$\mbox{cm}^{2}$}}

\begin{document}

\preprint{} \preprint{LYCEN 2007-14}
			 
\title{Brittle fracture down to femto-Joules --- and below}

\author{J.~{\AA}str\"om$^2$,
P.C.F.~Di Stefano$^{1*}$,
F.~Pr\"obst$^4$,
L.~Stodolsky$^{4}$,
J.~Timonen$^3$}

\affiliation{
$^1$  Institut de Physique Nucl\'eaire de Lyon, IPNL, UMR5822, CNRS-IN2P3; Universit\'e de Lyon, Universit\'e Claude Bernard Lyon 1, F-69622 Villeurbanne, France;\\
$^2$  CSC - IT Center for Science, P.O.Box 405, FIN-02101 Esbo, Finland;\\
$^3$ Department of Physics, P.O. Box 35 (YFL), FIN-40014 University of Jyv\"askyl\"a, Finland;\\
$^4$ Max-Planck-Institut f\"ur Physik, F\"ohringer Ring 6, D-80805 Munich, Germany;\\
$^*$ Corresponding author, {\it email address:}  distefano@ipnl.in2p3.fr.}

\date {\today}

\begin{abstract}
We analyze large sets of energy-release data created by stress-induced brittle fracture in a pure sapphire crystal at close to zero temperature where stochastic fluctuations are minimal.
The waiting-time distribution follows that observed for fracture in rock and for earthquakes.
Despite strong time correlations of the events and the presence of large-event precursors,  
simple prediction algorithms only succeed in a very weak probabilistic sense.
We also discuss prospects for further cryogenic experiments reaching close to single-bond sensitivity and able to investigate the existence of a transition-stress regime.
\end{abstract}


\maketitle
\begin{figure}[th]
\includegraphics[width=\textwidth]{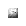}
\caption{%
Confocal microscopy picture of sapphire crystal fractured by sapphire bearing.  Diameter of affected area is $\approx 2 \mbox{ mm}$.  Slide marks are visible, as are irregular fractures of radial and circumferential type.
}
\label{distefano_Fig_Fracture}
\end{figure}
\begin{figure}[th]
\includegraphics[width=\textwidth]{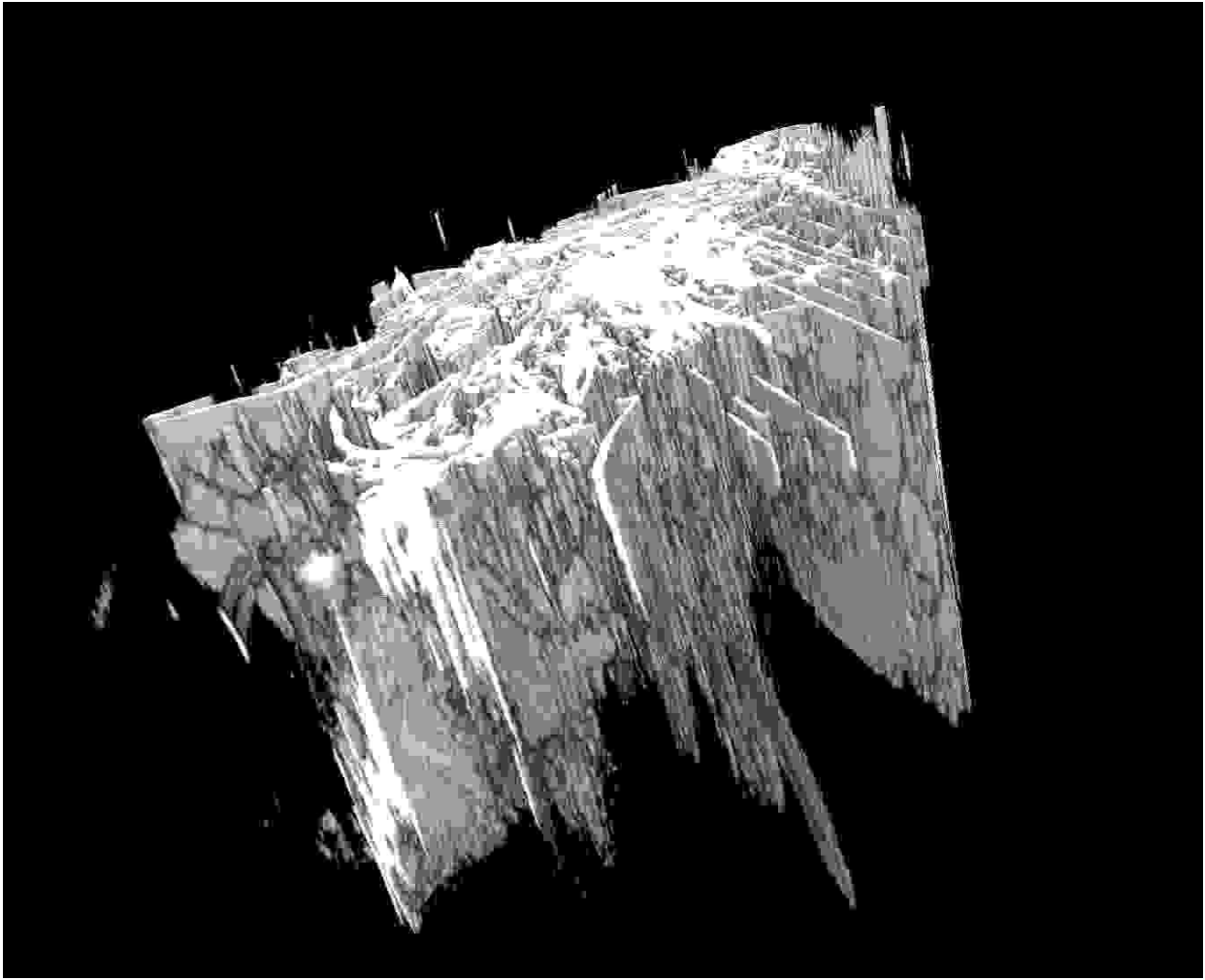}
\caption{%
Fractured area of the crystal scanned by confocal microscopy to a depth of about $150 \ \mu\mbox{m}$. Features visible at the sapphire surface are openings of two-dimensional cracks that extend fairly deep into the crystal. Their depth could not be measured accurately with the present method. This three-dimensional image was made by volume rendering of the cracks only, i.e. the solid material is not shown here.}
\label{distefano_Fig_Fracture_2}
\end{figure}

We have recently described a serendipitous and novel measurement of brittle fracture using cryogenic calorimetry~\cite{art:cracks_PLA_short}.  In a 260~g pure sapphire crystal cooled to 20~mK, cracks formed under pressure from sapphire bearings in what amounts to a sharp indentation experiment (Fig.~\ref{distefano_Fig_Fracture},~\ref{distefano_Fig_Fracture_2}).  
The small contact surfaces generated stress fields vanishing quickly with distance, and ensured stability of the fractures~\cite{book:Lawn_chpt8}.
The calorimetric measurement provided a direct measurement of the energy of the phonons from fracture events, and great sensitivity, 
of the order of a few femto-Joules.  
The rich and complete event catalogues, of many thousand femto-fractures each, contain the arrival time and energy of each event, and show several statistical similarities to earthquakes, despite the many orders of magnitude  difference in the energy ranges~\cite{art:Corral_PRL_2004}.  The similarities include:
({\it i}) the probability distribution of fracture-energy release is a power law with an exponent close to that of the differential Gutenberg-Richter relation expressed for seismic moment (which is proportional to earthquake energy)~\cite{book:Scholz}, 
({\it ii})
fracture events are long-range correlated in time with a power-law waiting-time distribution for short times,
({\it iii}) the fracture time series has the characteristics of fractal Gaussian intermittant noise,
and ({\it iv}) there is an elevated event rate right after large events and a power-law event rate decay.
More generally, the absence of trends in the data indicate that this represents a new example of steady-state slow brittle fracture, in an ordered system.  Up to now, such fracture has been linked to the disorder inherent in self-organized-criticality~\cite{art:Zapperi_Nature_1997,art:Garcimartin_PRL_1997,art:Salminen_PRL_2002}.

In the following, we show that the waiting-time distribution follows a general power-law exponential form observed in earthquakes and rock fracture, with the same power.
We demonstrate further correlations in the data and attempt to use them as predictors of the large, catastrophic fractures that should eventually occur.
Lastly, we  discuss a dedicated cryogenic  experiment to study these phenomena, down to 
energies close to those of single bonds in the crystal.  Such an experiment could also probe putative stress-dependent variations of the fracture rate, and investigate the existence of a transition-stress regime where the average energy release would vanish.

\section{Waiting times and clustering}
\begin{figure}[th]
\includegraphics[width=\textwidth]{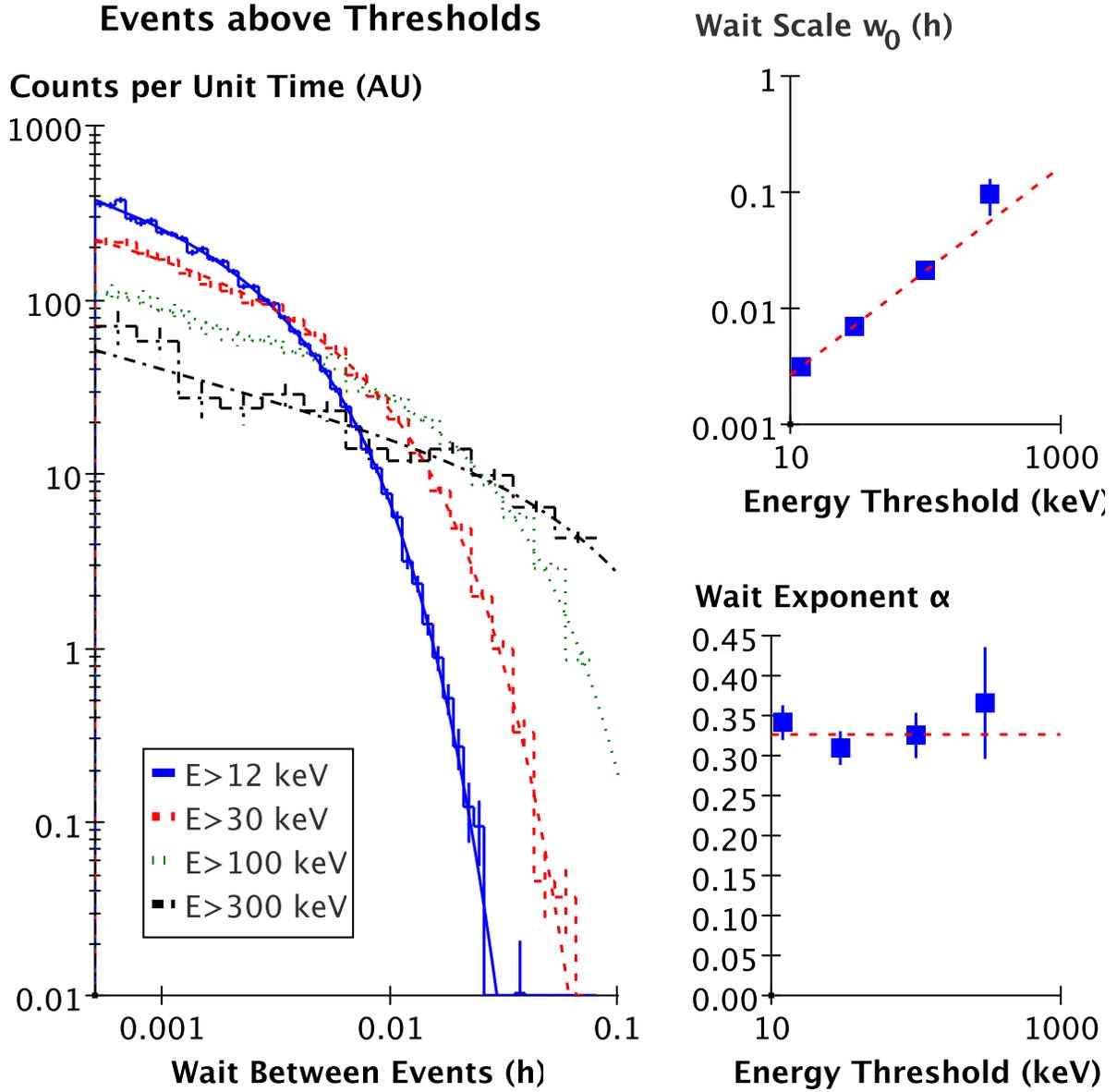}
\caption{Left: distributions of waiting times $w$ between consecutive events above various energy thresholds. Error bars are the square root of each bin content.
Data are well fitted by the product of an inverse power law and an exponential decay ($\propto \ w^{-\alpha} \exp{- w /w_0}$).
Top right: the wait distribution scale term, 
$w_0(\geq E)$, 
obtained from fits of the wait distribution (error bars are from fit) is compatible with fit by $\propto \  E^{\beta -1}$ (dashed red line), where $\beta = 1.9$ is obtained from the fit of the energy distribution~\cite{art:cracks_PLA_short}.  
Bottom right: wait exponent $\alpha$ obtained from fits, as a function of threshold energy (error bars are fit errors).  
Data are fitted by constant function yielding $\alpha = 0.33 \pm 0.01$.
These figures demonstrate that at least up to the highest energies, the wait scale term does indeed scale like $w_0(\geq E) \propto E^{\beta -1}$, whereas the wait power term does not depend strongly on the energy threshold.
}
\label{distefano_Fig_WaitTimes}
\end{figure}

We have shown~\cite{art:cracks_PLA_short, art:cracks_Comment_short} that the distribution of the waiting time between consecutive events above threshold, $w$, follows a power law at short times with an exponential fall-off at large waits: $dN/dw \propto w^{-\alpha} \exp{- w /w_0}$.  
The form of this waiting-time distribution is identical to that observed for earthquakes and rock fracture, 
further extending its validity~\cite{art:universal_cracks}.
Similar forms may be derived from the Gutenberg-Richter and Omori laws~\cite{art:Sornette_2006}.
The average wait must be proportional to the scale term $w_0$; integration by parts yields: $\overline{w}  = (1-\alpha) w_0$.
On the other hand, the distribution of events as a function of energy follows a power law: $dN/dE \propto E^{-\beta}$, with $\beta \approx 1.9$~\cite{art:cracks_PLA_short}.  By integration, the number of events above a given energy therefore also follows a power law: $N(\geq E) \propto E^{-\beta +1}$.  This is inversely proportional to the average waiting time for events above a threshold: $\overline{ w (\geq E)} =  (1 - \alpha) w_0(\geq E) \propto E^{\beta -1}$.  
Fig.~\ref{distefano_Fig_WaitTimes} shows that, as threshold energy increases,  the distribution of waiting times retains the form $w^{-\alpha} \exp{(- w /w_0(\geq E))}$, where the power $\alpha$ has little dependence on the threshold energy and $w_0$ scales like $E^{\beta -1}$.  
A fit of the wait power yields $\alpha = 0.33 \pm 0.01$.  With the notations from Ref.~\cite{art:universal_cracks}, we find $B= \frac{1}{1-\alpha} = 1.49 \pm 0.02$ and $\gamma = 1 -\alpha = 0.67 \pm 0.01$.  
These values are strikingly close to those obtained for earthquakes~\cite{art:Corral_PRL_2004} and are also compatible with those for acoustic emission of fractures in rocks~\cite{art:universal_cracks}.

If the energy distribution can be extrapolated to values large enough to cause a catastrophic destruction of the detector itself, then such a catastrophic event would arrive in a long, perhaps, but finite, time.  
If $E_{cat}$ is the energy released as the crystal breaks, then, in this setup,  it would be expected after $\overline{ w_{cat}} =   w_0(\geq E)  \left[E_{cat} / E \right]^{\beta -1} \approx ( 0.003 \mbox{ h}) \times \left[E_{cat} / (10 \mbox{ keV}) \right]^{0.9}$.  This scales slightly slower than a linear relation.  
To obtain an order of magnitude of the timescales involved, we assume the fracture surface energy of sapphire is an upper limit on the energy that would be released by a crack.  For instance, taking a value of $7.3 \mbox{ J/m}^2 \approx 4.5 \times 10^{12} \mbox{ keV/cm}^2$ for the surface energy of the $\left\{ \bar{1}012 \right\}$ plane~\cite{art:Wiederhorn_1969}, 
and defining a catastrophic crack size as $1 \mbox{ cm}^2$ in the $4 \times 4 \times 4 \mbox{ cm}^3$ cubic  crystal, 
this translates to an upper bound of $E_{cat} \leq 4.5 \times 10^{12} \mbox{ keV}$.
The weak upper limit on the expected wait for such a catastrophic event is therefore several millenia.

\begin{figure}[th]
\includegraphics[width=\textwidth]{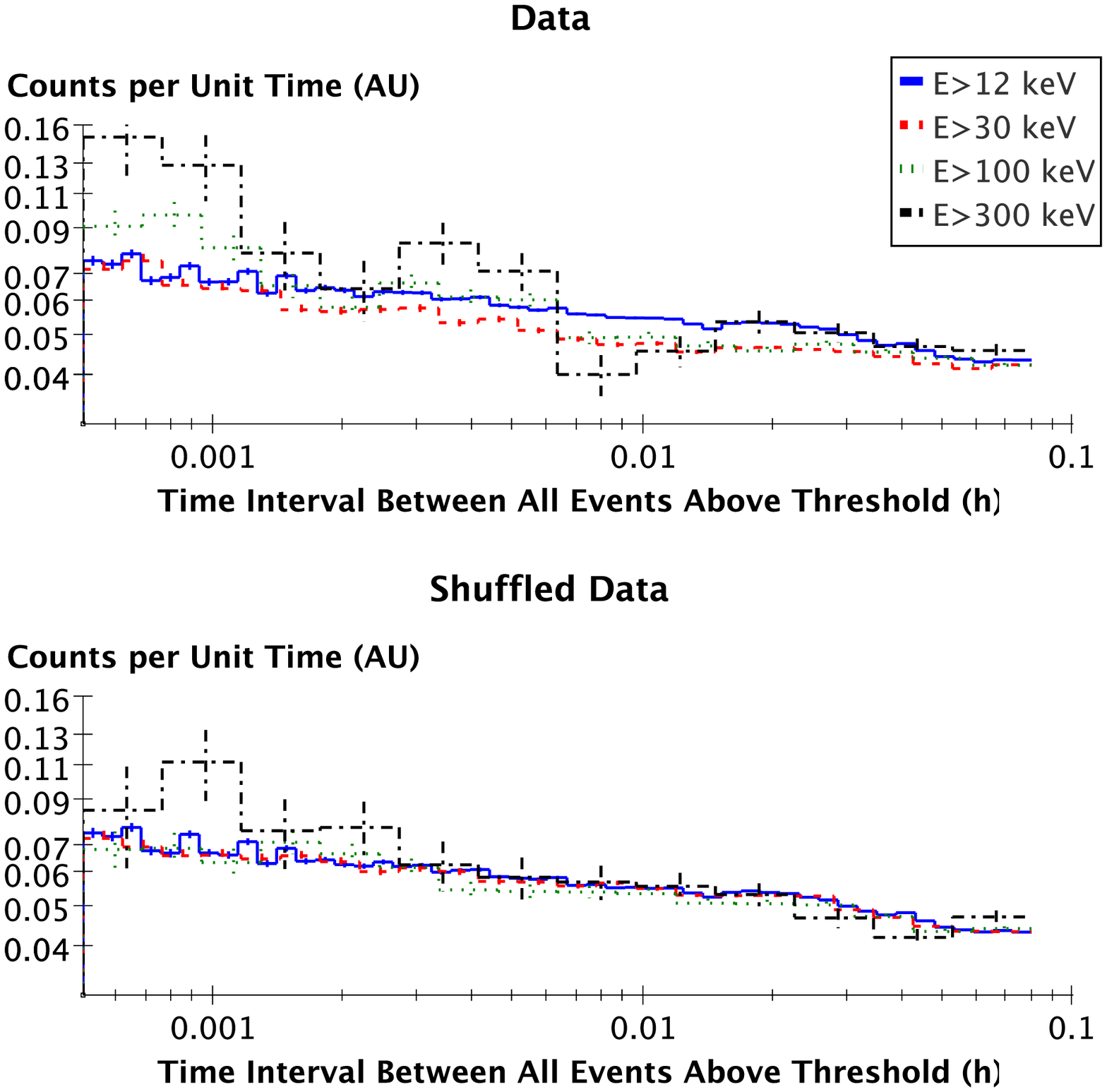}
\caption{Distribution of  times between all (as opposed to consecutive) events in various energy ranges (note that binning varies between ranges).  Top: in the data, the distribution becomes peaked at low times as the event size increases.  This is an indicator of clustering for large events. 
Bottom: in a random shuffle of the data, the distributions differ less for all event sizes.
}
\label{distefano_Fig_TimeDifference}
\end{figure}

The distribution of time intervals between all, rather than consecutive, events above various energies is shown in Fig.~\ref{distefano_Fig_TimeDifference}.  As energies increase, the distribution becomes more and more peaked at low time intervals. This is further indication that the large events cluster~\cite{art:cracks_PLA_short}.  As a control, the same analysis is applied to a random shuffle of the arrival times in the data.  In the shuffle, there is no energy-dependent effect.

\section{Weak predictability}
\begin{figure}[th]
\includegraphics[width=\textwidth]{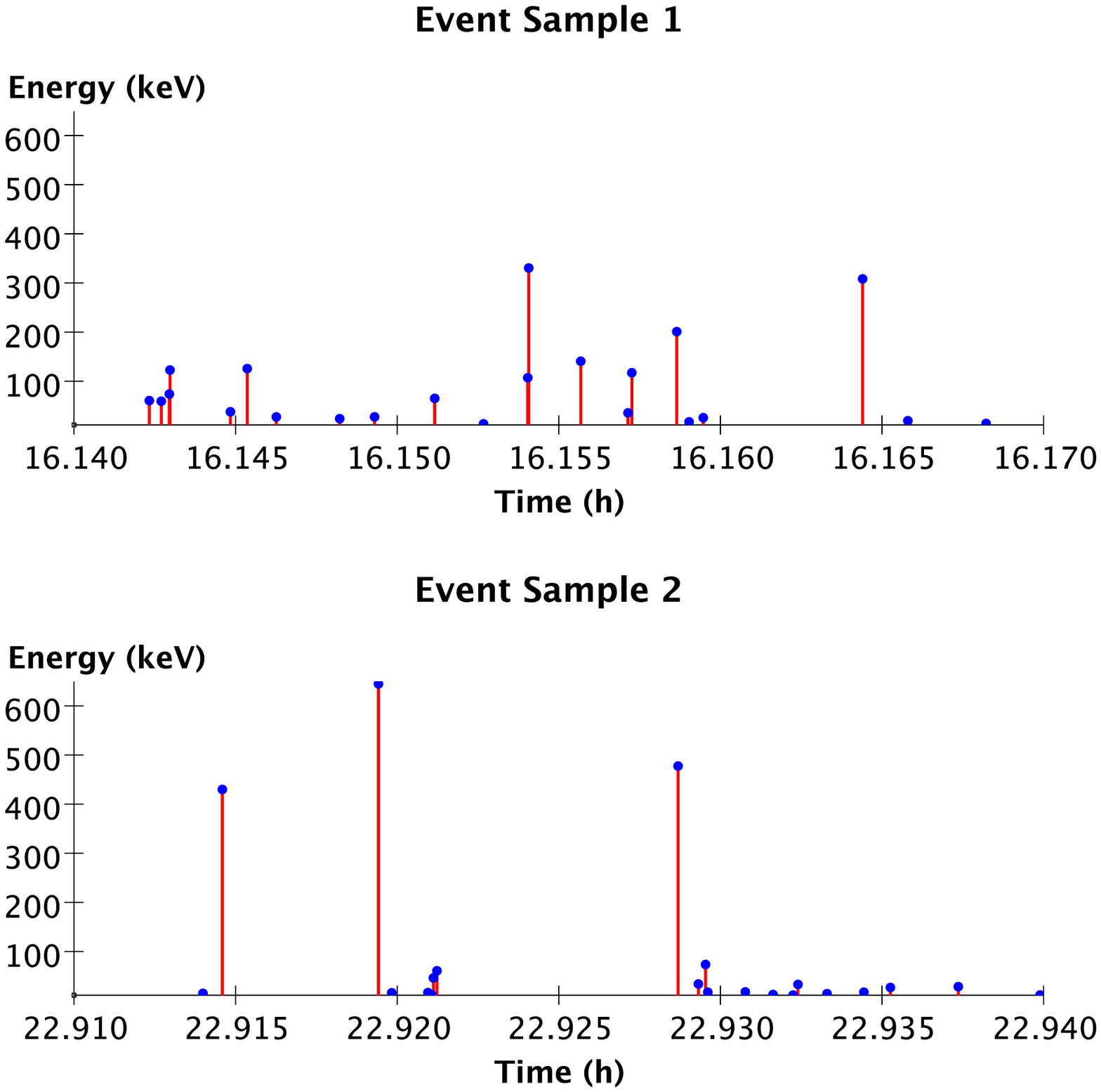}
\caption{%
Examples of event energies as a function of time.  Some large events ($E > 300 \mbox{ keV}$) appear after relatively quiet periods (for instance just before 22.92~h) whereas others have some precursors (for instance just before 16.155~h) and others display aftershocks (around 22.93~h)
}
\label{distefano_Fig_EvsT}
\end{figure}

\begin{figure}[th]
\includegraphics[width=\textwidth]{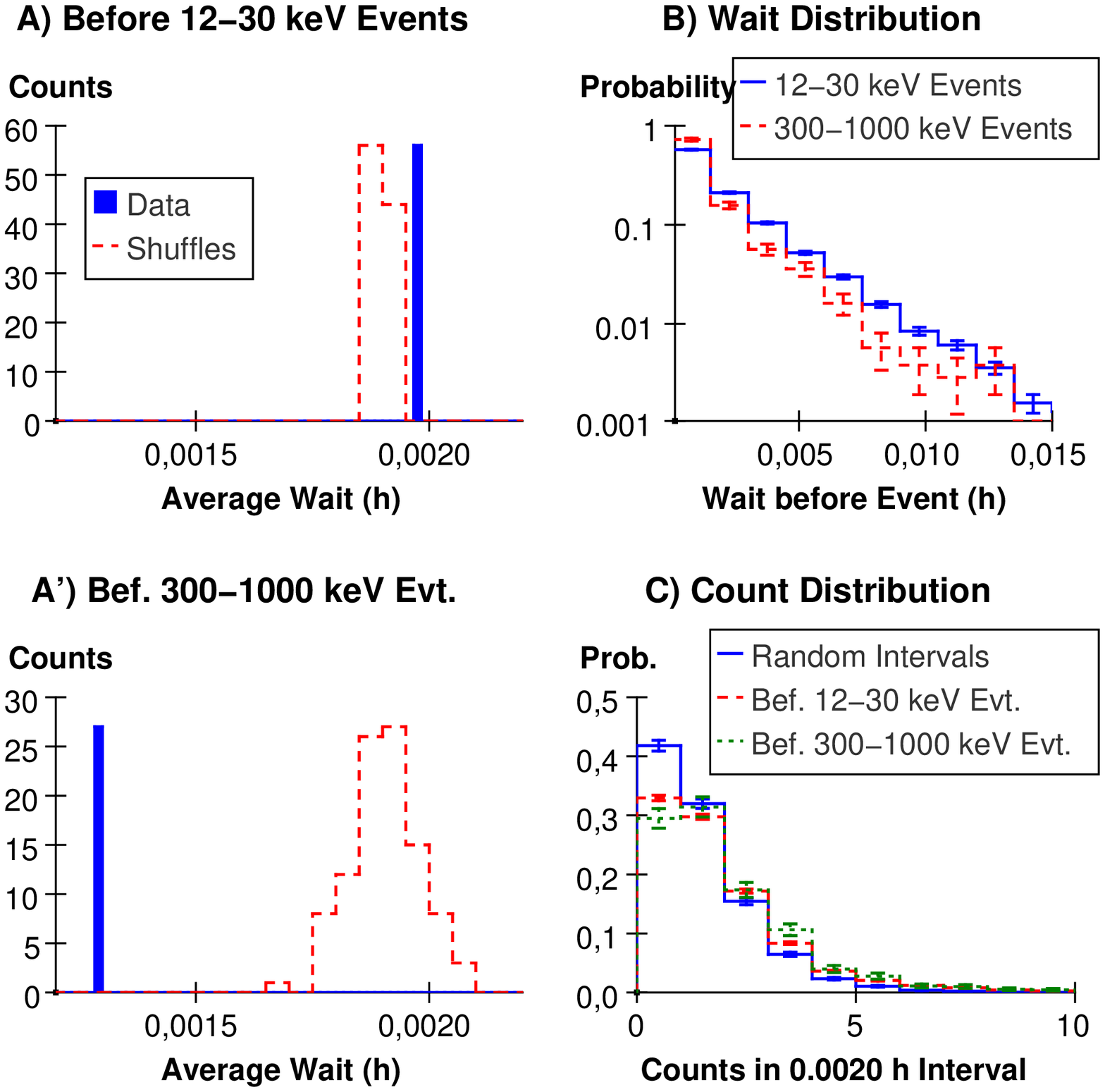}
\caption{
Fig. A shows the average wait between a small (12--30~keV) event and the event of any size preceeding it (solid blue bar), and the distribution of this average wait before small events in 100 shuffles of data (dashed red histogram).  
Fig. A' shows the same but for large events (300--1000~keV).  Fig. A and A' demonstrate that there is significantly less wait on average before a large event than before a small one.  
However, Fig. B illustrates that though the averages differ, the distribution of waiting times before small and large events are quite similar and will not provide strong discrimination between individual small and large events.  
Fig. C shows the distribution of counts in random intervals (solid blue line), in intervals before small events (dashed red) and in intervals before large events (dotted green).  The distributions do not show a strong distinction between small and large events, and indeed little difference between random intervals and intervals preceeding events.}
\label{distefano_Fig_Mosaic}
\end{figure}

Some examples of event-energy time series from a 50~h run  containing $\approx 30000$~events above a threshold of 12~keV are shown in Fig.~\ref{distefano_Fig_EvsT}.  Various cases are visible, including a large event with a precursor ($t \approx 16.154 \mbox{ h}$),  a relatively isolated large event ($t \approx 22.915 \mbox{ h}$), and a lull before a large event followed by aftershocks ($t \approx 22.93 \mbox{ h}$).
%
%
%
In Figures~\ref{distefano_Fig_Mosaic}~A and A', we plot the average value of the waiting time before each of the $\approx 14000$ small events (12--30~keV) and $\approx 1100$ large events (300-1000~keV).  
The waiting times for small (respectively large) events are here defined as the wait between a small (resp. large) event and the preceeding event regardless of its size.
On average, there is less wait before large events (0.0013~h) than before small events (0.002~h).  To check if this a statistical fluctuation, we generate 100 shuffles of the data set, by randomly permuting the arrival times of the events, then do the same analysis as on the original data.  The distribution of the average values of the shuffles does not cover the spread of the real values, confirming that in the actual data, the wait is shorter before large events than before small ones.  This appears to be another manifestation of the increase in rate which is particularly evident around large events~\cite{art:cracks_PLA_short}.

These and the numerous other correlations present in the data provide motivation to attempt prediction of large events, a  challenge  of relevance for other phenomena, ranging from avalanches in snow~\cite{art:aastrom_pre2001}
to earthquakes~\cite{book:Scholz}. 
Fig.~\ref{distefano_Fig_Mosaic}~B  compares the distribution of waiting times before small and large events.   The significant correlations present on average are much harder to exploit on an event-per-event basis, as the distribution for large events does not differ greatly from that for small events.

We also attempt to predict the arrival of large events using the distribution of events in a given time window  (Fig.~\ref{distefano_Fig_Mosaic}~C).  Window duration is 0.002~h, corresponding to the average waiting time in the run.  For comparison, we generate 5000 random intervals. The difference between the distribution of counts in the random intervals and in the intervals preceeding small or large events is slim, while the difference between intervals preceeding small and large events is slighter yet.  
With these simple methods, predictability of individual large events is therefore poor.

While the weak predictability we have described here could perhaps be enhanced by more sophisticated algorithms, it might also be either a general conclusion for brittle fracture, or indicate that, in our particular setup, the combination of several crack systems propagating independently masks any individual patterns and predictability.

\section{Prospects for further study}
%
%
Further study would benefit from a dedicated cryogenic detector with only a single bearing creating the cracks.  
It could allow investigation of fractures down to low energies.
Additionally, if fracture rate is found to depend on applied stress, 
it might allow investigation of low rates close to a possible transition stress at which fractures just appear.
If such a regime exists, the waiting-time distribution would be a pure power-law, and the whole system could be in a critical transition at which the average energy release rate vanishes.

\begin{figure}[th]
\includegraphics[width=\textwidth]{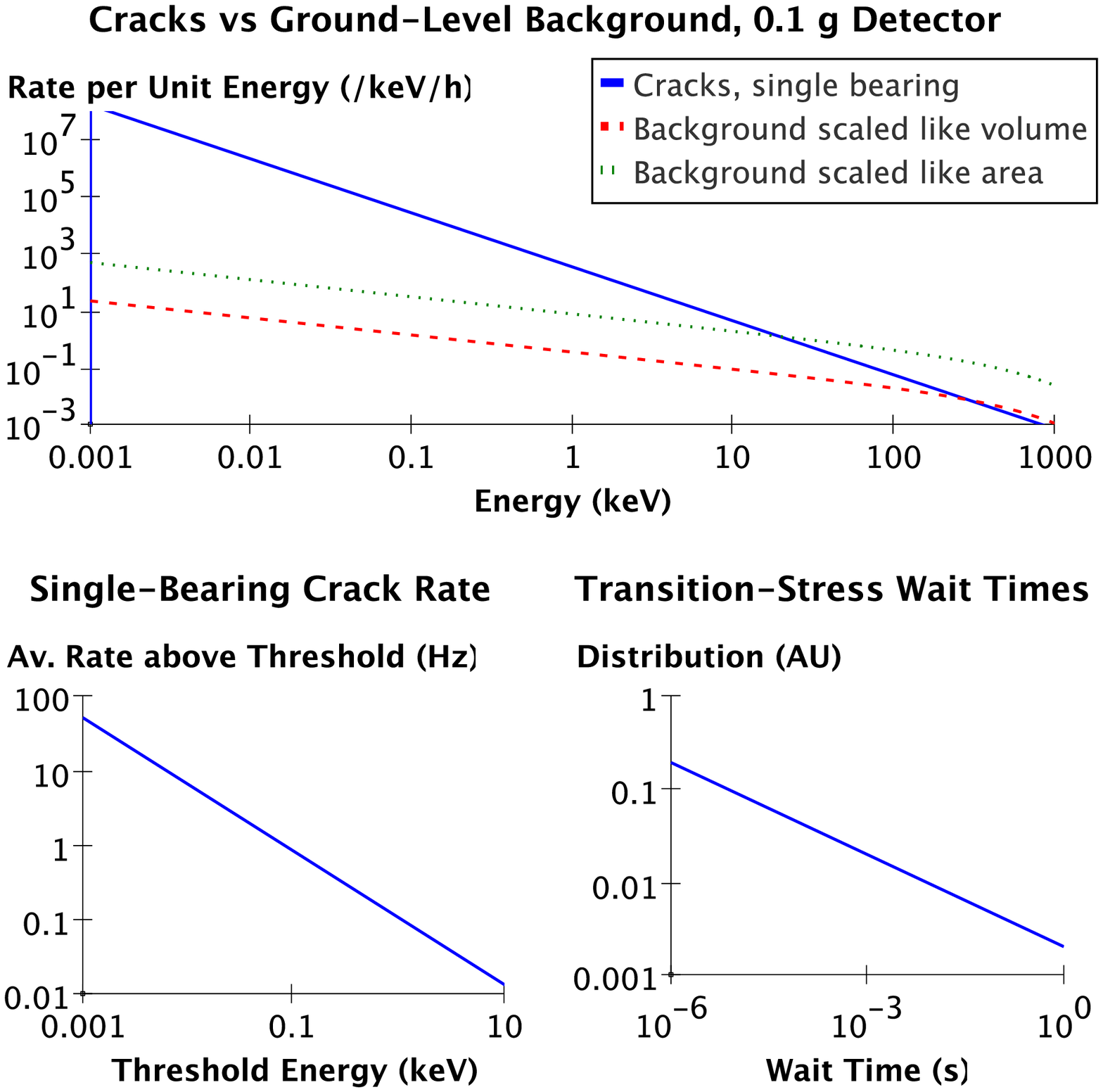}
\caption{%
Top: crack rate compared to backgrounds in hypothetical 0.1~g detector in the case of a background scaling like detector mass and in the case of one scaling like detector area.  
Rate of cracks is scaled to a single pressure point, 
though in a new experiment this rate could perhaps be fine-tuned between cryogenic runs by altering the force applied on the bearing.  Backgrounds are natural radioactivity and cosmic-induced at ground level, and are extrapolated from Fig.~2 of Ref.~\cite{art:ROSEBUD_1kg} to low energies and to low detector masses.  Being able to observe the cracks at ground level will require a  small crystal and perhaps an increased rate of cracks. 
Bottom left: the crack rate integrated above even a low threshold remains, on average, compatible with typical thermal time constants of cryogenic detectors and typical acquisition rates.
Bottom right: the distribution of waiting times, here in the $\propto w^{-\alpha}$  transition-stress limit with $\alpha = 0.33$, shows that the abundance of short waiting times makes pile-up inevitable.}
\label{distefano_Fig_Background}
\end{figure}

One option would be to carry on in a low-background, deep-underground setup, such as that of the CRESST~II experiment~\cite{art:CRESST_AstroPart_2004}.  An existing detector holder could be modified  to include a single bearing pressing against one end of the cylindrical crystal of  40~mm height and 40~mm diameter.
In itself, this will require some ingenuity as the crystal itself must not move because of the bearing but cannot,  for thermal reasons,  be held firmly by large contact areas.
One of the currently standard \CaW\ crystals could be used, or a new \alo\ one could be manufactured.  
In either case, it would be interesting to retain the light detector of the CRESST~II setup to see if crack formation is accompanied by light emission, since fracto-emission of photons and electrons has been reported in other crystals~\cite{art:Langford_1987}.
Adjusting the tightness of the spring pressing the bearing to probe an effect on crack rate, and to reach transition stress if it exists, would have to be done between cryogenic cycles and would require some trial and error.
As in the original work, energy calibration would be obtained by an external, removable, \CoF\ source, providing 122~keV photons.  Lower-energy calibration would be obtained via heater pulses.  It should be possible to lower the phonon threshold down to around 1~keV compared to the $\approx$~10~keV in this work, though this gain would probably not be significant from the standpoint of brittle fracture.
Another underground setup that could be of interest is that of the EDELWEISS experiment which uses germanium ionization-phonon detectors~\cite{art:DiStefano_2000}.  In this case, there would be a simultaneous measurement of the phonons created by the cracks as well as of whatever ionization  the cracks create~\cite{art:Langford_1987}.  
Though the original work was carried out in a special low-background environment underground, it would be simpler if in the future it could be done in a standard cryostat on the surface.  
For this, the rate of crack events must be much larger than the rate of other events; this means that though a low-threshold experiment may be feasible, a transition-stress one will be difficult.
The rate of crack events may depend on the force pressing the bearing against the crystal but does not depend on the mass of the crystal itself.  
We assume that in a dedicated experiment, 
there would be only a single bearing rather than a dozen as in the original work, and therefore divide the original rate by 12.  
The competing backgrounds, mainly cosmic-ray-induced particles and radioactivity of the detector and its surroundings, both increase with detector mass.  As illustration, we extract a rough estimate of the sum of these backgrounds from previously published data for a partially shielded 1~kg sapphire detector at the Earth's surface~\cite{art:ROSEBUD_1kg}.  Between the threshold of 50~keV and $\approx$~1~MeV, the background follows the product of an inverse power law, with an exponent of 0.6, and a decaying exponential, with a typical energy of $\approx$~600~keV, high above the range relevant here.
In Fig.~\ref{distefano_Fig_Background}, we compare the scaled rate of cracks to the  backgrounds in an 0.1~g sapphire detector for two different scaling laws of the background as a function of detector mass $m$: scaling proportional to mass (i.e. volume, $\propto m$), and scaling proportional to surface area ($\propto m^{2/3}$).  In both cases, we assume that the background power-law holds below 50~keV.  A significant background proportional to the surface area comes from cosmic muons of which there are $\approx 1 \mbox{ cm}^{-2} \mbox{ min}^{-1}$ depositing $\approx 500$~keV per mm of \alo\ passed through~\cite{PDBook_2006_short}.
Working at ground-level will require some combination of small crystals, low threshold, and, if possible, increased crack rate, 
though this last point is incompatible with a reduction in stress to the level at which fractures may nearly vanish.  
Energy calibration of a small crystal requires a low-energy radioactive source such as \FeF; the source  must be placed within the cryostat for calibration, though it may be possible to remove it or block it during data-taking for fractures.
Such a small crystal could have a threshold of less than a few hundred~eV, and as energy decreases, the crack rate ($\propto E^{-1.9}$) rises faster than the extrapolated background ($\propto E^{-0.6}$).
%

%
In addition, an energy  threshold below a few~eV could make the device sensitive to the rupture of single sapphire bonds.
A binding energy of 7.34~eV per atom has been reported for \alo~\cite{art:Hong_2005}, 
though it is not apparent to us what phonon energy accompanies rupture of a  bond.
A 10~eV threshold is achievable given current, $\approx \cmsq$, cryogenic detector development.  
For instance, the CRESST experiment has developed thin silicon calorimeters, of surface area several $\mbox{cm}^2$ and thickness about $0.5 \mbox{ mm}$, with thresholds better than 40~eV~\cite{art:distefano2003_short,art:Petricca_LTD10}.
These detectors are optimized for light detection  rather than for a low threshold per se.
A smaller, parallelipiped-shaped, \alo\ device, measuring $5 \times 5 \times 1 \mbox{ mm}^3$ (m=0.1~g), with an optimized thermometer,  should be able to reach lower thresholds, while remaining large enough for a pressure-bearing $1 \mbox{ mm}$ diameter sapphire or diamond ball.  
The device could be mounted  $\approx 1 \mbox{ mm}$ from a CRESST-type light detector of similar size, to see any light produced by the fractures~\cite{art:Langford_1987}.
We note that to obtain an absolute energy calibration at 6~keV from \FeF\ and to have a threshold of 10~eV will require a dynamic range of about three orders of magnitude which may be difficult to obtain with a transition-edge sensor.  Another challenge will come from the rate of cracks.  Detectors of size $3 \times 3 \times 0.5 \mbox{ mm}^3$, optimized for speed rather than threshold, reach rise times of $\approx 1~\mu \mbox{s}$~\cite{art:Rutzinger_LTD10_short} (smaller devices in which the transition-edge sensor itself is the absorber can be an order of magnitude faster~\cite{art:Cabrera_PB2000}).  
Though this may be compatible with the average rate of cracks, pileup will be inevitable, even for an arbitrarily fast detector, since the distribution of waiting times contains an inverse-power-law term (Fig.~\ref{distefano_Fig_Background}).

The calorimetric technique is readily applicable to many other dielectric materials, such as \CaW, Ge and Si already mentioned. As we have stated, we are not however aware of a clear relationship between the elastic energy used to break bonds and the elastic energy left over in phonons which we measure.  
Nonetheless, the partition of energy is simpler than in the case of acoustic emission, where only a fraction of phonons are measured.  The calorimetric technique could therefore provide new insight into the mechanics of fracture.

\section{Conclusion}
The distribution of waiting times between brittle fracture events observed in a cryogenic detector contains a power-law term which is independent of energy threshold and an exponential scale that depends on it.  This form matches that previously observed by acoustic emission in rock and that  observed for earthquakes.
If the energy distribution holds for events large enough to shatter the detector, then such an event is expected in a very long, but finite amount of time.
Though we have shown additional correlations in the data, predicting such large, catastrophic, fractures is not straightforward.  
To see if this is due to the multiple sources of cracks in this data, we propose a dedicated experiment with a single pressure point.  With typical fracture rates observed heretofore, such an experiment is feasible at ground level with a smaller cryogenic detector if extrapolation of background holds to low energies and masses.  
However, searching for vanishing fracture rates  requires at least a shallow underground site.  
In either case, the lower threshold associated with a smaller, optimized, detector would enable it to probe brittle fracture down close to the energy of single bonds in the crystal.  The calorimetric technique could provide additional insight into the partition of elastic energy into permanent dislocations and phonons.

\section{Acknowledgments}
P.~Di~Stefano acknowledges discussions with M.~P.~Marder on the
physics of brittle fracture.  I.~Maasilta has provided helpful comments on low-threshold detectors.
The Java implementation of the Abstract Interface for Data Analysis (\url{http://java.freehep.org}) has been used for data analysis and figure preparation.

\bibliography{AbreviationsShort_01,Calorimeters,DMandDD_01,Fractures,ParticlePhysics_01,Seismology,Scintillation,SolidState_01}

\newpage

\newpage

\newpage

\newpage

\newpage

\newpage

\newpage

\end{document}